%% file: main.tex
\documentclass[twocolumn, aps, pra, amsmath, amssymb, nofootinbib, groupedaddress, longbibliography, floatfix, table-of-contents, eqsecnum, dblfloatfix]{revtex4-2}

\usepackage[pdftex]{graphicx}
\usepackage{float}
\usepackage{mathrsfs}
\usepackage[colorlinks, breaklinks, urlcolor={blue}, linkcolor={red}, citecolor={blue}]{hyperref}
\usepackage{array}
\usepackage{amsmath}
\usepackage{type1cm}
\usepackage{lettrine}
\usepackage[english]{babel}
\usepackage{lmodern}
\usepackage{microtype}
\usepackage{booktabs}
\usepackage[T1]{fontenc}
\usepackage{caption}
\usepackage{braket}
\usepackage{todonotes}

\setuptodonotes{size=\tiny}

\begin{document}

\title{Pushing Boundaries: Quantum-Enhanced Leader Election and the Limits of Consensus}

\author{Chandrashekar Radhakrishnan}
\email[]{cr2442@nyu.edu}

\author{Yuhang Zheng}
\email[]{yz6209@nyu.edu}

\author{Olivier Marin}
\email[]{ogm2@nyu.edu}

\affiliation{Department of Computer Science and Engineering, New York University Shanghai, 567 West Yangsi Road, Pudong, Shanghai 200124, China}

\begin{abstract}
This work addresses the complexities involved in designing distributed quantum algorithms, highlighting that quantum entanglement does not bypass 
the Fischer-Lynch-Paterson (FLP) impossibility theorem in asynchronous networks. Although quantum resources such as entanglement offer potential 
speedups, the inherent constraints of classical communication remain. We develop a leader election algorithm as a proof of concept, demonstrating 
how entanglement can enhance efficiency while still contending with asynchronous delays. This algorithm serves as a foundation for a broader 
blueprint for future distributed quantum algorithms, providing insights into both the real performance gains and the limitations that entanglement 
offers in a distributed setting. 
\end{abstract}

\maketitle

%
%
%

\section{Introduction}
Reducing the processing time for hard problems is one of the most critical issues for computing. 
This is achievable in two complementary ways: by designing efficient algorithms with better time 
complexity, and by increasing the computing power of the system that runs the algorithms. 
This article focuses on the latter in the context of quantum computing.

Quantum computing increases computational power through quantum advantage: in comparison with a 
traditional computer, it offers a considerable speedup of processing time for selected problems. 
Quantum computing is now a reality, with the advent of practical quantum computers based on different 
technological platforms like superconducting systems \cite{barends2013coherent}, trapped ions \cite{cirac1995quantum} 
and photonic systems \cite{o2009photonic}. The efficiency of these quantum computers is evolving fast, with the expectation that they
will become commercially useful in the near future. As happened in the past with classical 
computers, however, a quantum computer is perceived as a stand-alone unit for the centralized
execution of an algorithm. 

Another way of increasing computational power is to distribute computations over multiple processors. 
Every processor handles an independent portion of the computation concurrently, and shares its partial 
results with all the other processors so they can form the complete results together. Remote processors 
must reach consensus in order to achieve a consistent view of results, for instance to agree on the value of 
shared data, or to dispatch tasks among processors.

Distributed quantum computing is a combination of both paradigms: a set of quantum computers 
cooperate over a network to increase their common computing power.
The literature offers several distributed versions of quantum algorithms~\cite{dqc-survey, dqc-consensus-closer-look, quantum-consensus-blockchain, distr-grover}. However, all of these 
remain theoretical; they overlook critical distributed computing constraints. Chief among these is 
that exchanges of data between processors introduce delays, which are an obstacle to consensus. 

This work contributes two significant results that address these challenges at the intersection of 
distributed and quantum computing. 
It addresses the complexities of designing distributed quantum algorithms, demonstrating that quantum entanglement does not circumvent the limitations posed by the FLP impossibility in asynchronous networks. As a proof of concept, we develop a quantum-enhanced leader election algorithm that leverages GHZ (Greenberger–Horne–Zeilinger) states to improve efficiency while factoring in communication delays. This algorithm serves as a foundation for a blueprint, which we present for future distributed quantum algorithms, outlining practical design principles that account for asynchrony, classical communication constraints, and entanglement advantages. Our findings illustrate both the potential gains and the limitations of entanglement in distributed quantum systems, providing insights into the conditions for efficient operation and the enduring challenges of achieving consensus in quantum-enhanced architectures.

The remainder of this paper is organized as follows. 
Section~\ref{sec:overview} provides an overview of foundational concepts in distributed systems, setting the stage for the challenges of distributed quantum computing. 
Section~\ref{sec:classical leader election} examines classical leader election algorithms to illustrate the constraints that distributed quantum algorithms must address. 
Section~\ref{sec:distr-quantum-fw} defines the hardware and operational requirements for integrating quantum components in a classical network. 
Section~\ref{sec:distr-quantum-election} introduces our quantum-enhanced leader election algorithm, describing its use of GHZ states to improve efficiency while factoring in communication delays. 
In Section~\ref{sec:blueprint}, we provide a blueprint for designing future distributed quantum algorithms. 
Finally, Section~\ref{sec:conclusion} summarizes our contributions and draws potential directions for future research.

%
%
%

\section{Distributed systems - an overview} \label{sec:overview}

To understand the challenges and potential of distributed quantum computing, we begin by reviewing 
the foundational principles of distributed systems. 

A computing device executes a sequential set of operations to solve a problem. Any given 
operation in a sequence need not be a single evaluation. In fact in many cases an operation 
contains multiple evaluations which can be performed simultaneously. However, a processor can 
only execute a limited number of evaluations at a time.  

Parallel and distributed systems speed up the whole computation by bringing together a collection 
of processors that carry out concurrent evaluations.
In \emph{parallel systems}, all processors store the computational data in a common memory. Sharing at 
a single location guarantees data consistency. However, the shared memory becomes a bottleneck
as the number of parallel processors increases. This limits the scalability of the system.

\emph{Distributed systems} do not rely on shared memory. Every computing node manages its own local data,
and communicates with other nodes via messages over a network. This removes the shared memory bottleneck, at the 
cost of network overhead to ensure that all nodes maintain a consistent view of the computational data.
\emph{Consensus} is the fundamental problem in which multiple processors must agree on a single data value to 
ensure consistency across a network. Many algorithms, mostly based on~\cite{paxos}, offer to solve consensus.
They vary in terms of the assumptions they make about the system model and about the failure model. 
Subsection~\ref{subsec:distr-syst-model} covers the main aspects of how to model a distributed system.

\subsection{Modeling the distributed system} \label{subsec:distr-syst-model}
The formal representation of a distributed system typically comprises three aspects: the logical organization of 
the network, the level of resistance to expect if failures occur, and the time it takes to perform local operations 
and to transfer messages.

\subsubsection{Topology}
In a distributed system each processor is considered as a node connected to other nodes 
through communication channels.  Considering the nodes as vertices and the communication 
channels as edges, one can obtain a graph-theoretic representation of the distributed 
system.  The graph representation provides the following properties characterizing
the distributed system:
\begin{itemize}
    \item \emph{Unidirectional/bidirectional.} In the graph theoretic representation, depending on whether the message is exchanged in one 
direction or in both directions, the distributed system can be denoted by a unidirectional 
or bidirectional graph.  Hence a distributed system can have a unidirectional or bidirectional
connectedness.
    \item \emph{Diameter of the network.} The diameter of a network is the maximum of the shortest path between two nodes.
    \item \emph{Degree of a node.} The degree of a node is the number of nodes it shares an edge with. These nodes are called \emph{neighbors}.
\end{itemize}

The system topology is crucial in that it influences the algorithmic design, and determines its potential efficiency.
Some special topologies include unidirectional rings, stars, spanning trees, and cliques. 

\subsubsection{Failure model} 
A distributed system involves multiple interconnected components that communicate and 
coordinate tasks. The more components there are, the higher the likelihood that one of them will fail at some point. In this context, a failure refers to any event or condition where a system 
component, either a node or a communication link, does not achieve its intended purpose. 
The outcome of the component's misbehavior allows to classify failure semantics~\cite{understanding-ft-disys} as {\it(a)} crash failures, {\it(b)} omission failures, {\it(c)} timing failures, and {\it(d)} arbitrary failures. 
A component that stops functioning altogether constitutes a crash failure.
An omission failure is a message exchange that doesn't proceed. Three different kinds of faults produce
such a failure: the source doesn't send the message, the destination doesn't receive it, or the link doesn't carry it. 
A timing failure is a node or a network operation that does not complete within its specified time 
bounds, such as a timeout, or a delayed response.
An arbitrary failure occurs when a system component exhibits a non-deterministic behavior unpredictably, 
possibly providing incorrect results or unauthorized actions due to bugs or malicious activity.
Determining failure semantics is one of the aspects of modelling a distributed system.

\subsubsection{Temporal model}
One of the crucial factors for successful communication between the nodes lies in handling message transfers 
in a time-sensitive manner.  Based on temporal properties, the distributed system can be modeled as:

{\it (a) Synchronous:}  
if there is a \emph{known} upper bound on message transmission delays.  

{\it (b) Partially Synchronous:}  
if there \emph{exists} an upper bound on message transmission delays, but there's no way to determine its exact value. 

{\it (c) Asynchronous:} 
if message transmission delays are unbounded; communications may require an infinite amount of time to proceed, which therefore includes failures.

Real physical networks are asynchronous systems by nature.

\subsection{Distributed Algorithms}
Once the system model is set, the next step is to investigate algorithms specifically built
for it.  Distributed algorithms must satisfy two important properties {\it viz} 
{\it (i)} safety and {\it (ii)} liveness.  The term safety refers to bad things that should 
not happen in any execution.  The liveness property specifies the good things that must 
happen eventually in every execution.  Assessing the performance of distributed algorithms 
usually relies on two classes of complexity: 

{\it (a) Message complexity:} 
Message complexity depends on two factors namely {\it(i)} Total number of messages 
exchanged and {\it (ii)} Total size of the exchanged data. 

{\it (b) Time/round complexity:} 
Time complexity is defined as the longest chain of messages where the unit of time is 
defined as the exchange of a message between two nodes. It differs from message complexity
because messages can be sent concurrently. Time complexity measures the causal sequence of 
message transfers required to reach algorithm termination.

\subsection{The FLP Impossibility Theorem}

This theorem states that it is impossible for an asynchronous system to achieve \emph{consensus}
in a deterministic way if even a single process crashes. In distributed systems, consensus 
is a fundamental problem, as multiple nodes in a network must agree on a shared value to 
ensure system consistency. Fischer, Lynch, \& Paterson proved that deterministic consensus 
is unattainable under normal networking conditions~\cite{flp-impossibility}. The intuition 
is that, due to potential communication delays of unknown duration, it is impossible to 
sort out computing nodes and communication links that are slow from those that have crashed altogether. 
This finding has spurred the development of a variety of consensus and coordination algorithms under more 
restrictive or probabilistic conditions, including models that assume partial 
synchrony~\cite{partial-synch}. In a partially synchronous distributed system, network delays and processing speeds may vary unpredictably, but will eventually meet known bounds, allowing the system to temporarily operate synchronously.

%
%
%

\section{Leader election algorithms in classical distributed systems} 
\label{sec:classical leader election}

\input{classic-election}

\section{A Framework for Distributed Quantum Systems}
\label{sec:distr-quantum-fw}

\input{dqc-framework}

\section{Quantum Leader Election Algorithm}
\label{sec:distr-quantum-election}

\input{quantum-election-algo}

\section{Blueprint for Designing Distributed Quantum Algorithms}
\label{sec:blueprint}

\input{blueprint}

\section{Conclusion}
\label{sec:conclusion}

This paper addresses the challenges associated with designing distributed quantum algorithms. In particular, it emphasizes that quantum entanglement does not overcome the FLP impossibility in asynchronous systems. We developed a quantum-enhanced leader election algorithm as a proof of concept. Our algorithm uses GHZ states to increase efficiency while accounting for the delays inherent in classical communication channels. This algorithm forms the basis of a blueprint we present for future distributed quantum algorithms. Our blueprint establishes design principles that integrate quantum resources with the realities of asynchronous networks. Our findings underscore both the potential and limitations of quantum entanglement in distributed settings, providing a practical framework for the scalable deployment of quantum resources in networked environments.

This work lays foundational steps toward establishing quantum computing as a scalable, practical technology. Several directions could further expand on our results. First, other well-established distributed protocols merit exploration to address timing and coordination issues in quantum algorithms on real networks. Such research could lead to more resilient and scalable quantum systems. While we focused on leader election for simplicity, more complex consensus protocols would be a logical next step, with leader election as a powerful starting point. Probabilistic approaches also present promising avenues for achieving consensus under quantum constraints. Finally, given the limited availability of quantum resources, optimizing their allocation across nodes will be crucial. Developing resource-efficient architectures could further enhance scalability and performance in distributed quantum networks.


%
\bibliography{refs}

\end{document}

%% file: classic-election.tex
In distributed systems, consensus is a fundamental problem requiring nodes to agree on a shared state, often in the presence of faults. Leader election, a simpler variant of consensus, requires nodes to agree on a single leader to coordinate actions or tasks. By focusing on leader election, this paper addresses the quantum-specific challenges in distributed settings without the full complexity of consensus. Leader election provides a concrete way to explore the limits and possibilities of quantum resources for distributed coordination.

In a leader election, nodes must agree on a single leader to coordinate distributed tasks, manage resources, handle consistency, and optimize system performance. 
Classic leader election algorithms operate by exchanging messages among nodes to agree on a leader, a process sensitive to network delays and failures.

Depending on the network’s fault tolerance requirements and structure, leader election algorithms can vary widely. This section categorizes these algorithms based on the constraints they impose on the system model: bounded delays on communications, the existence of unique node identifiers, the organization of nodes along a specific topology, and the tolerance to failures. Algorithms that assume a very constrained model are more likely to be simpler and more efficient.

\subsection{Ring Topology with Identifiers}

Leader election algorithms without fault tolerance target networks where nodes are reliable and communication is assumed to be uninterrupted. These algorithms typically operate in a known network topology and depend on nodes having unique identifiers. Most algorithms in this category are variations of Chang and Roberts’ algorithm, which we describe below.

\begin{figure}
    \centering
    \includegraphics[width=0.2\textwidth]{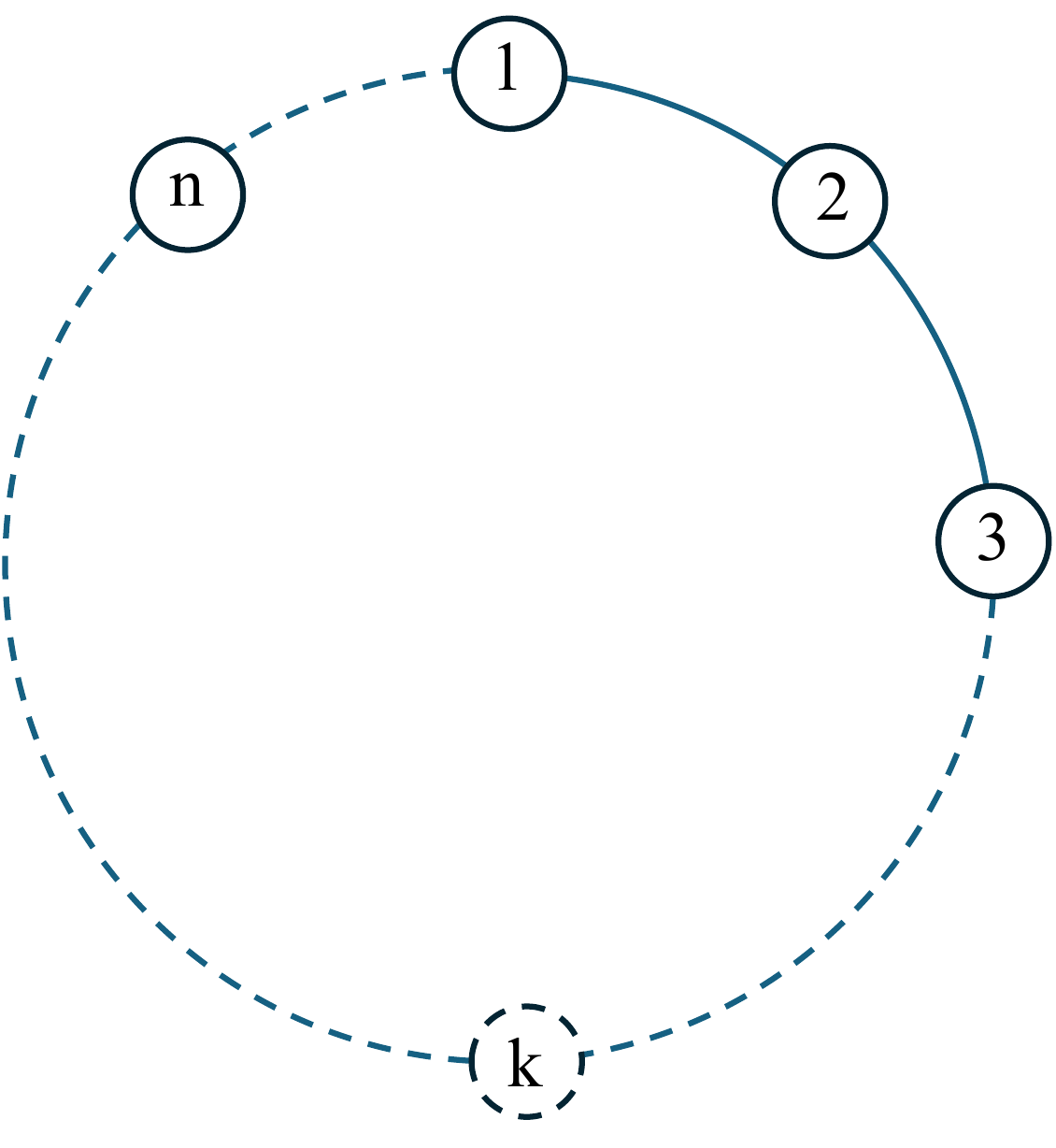}
    \caption{Distributed system in a ring topology}
    \label{fig:ring}
\end{figure}

Let us first consider a simple topology: the ring, as depicted in Figure~\ref{fig:ring}. In a ring topology, messages circulate through the network sequentially. Note that a logical ring structure can be created on any connected physical network -- one whose graph representation shows paths to reach any node from any other node.

Chang and Roberts’ algorithm~\cite{ring-elect-CR} elects a leader by having each node initiate an election when it doesn’t detect a leader. A node sends 
its ID via a message circulated around the ring. Each receiving node forwards the message if its ID is lower,  or replaces the ID if its own ID is 
higher. When a node sees its own ID returned, it declares itself the leader. With $O(N^2)$ messages required in the worst case, this algorithm is simple 
but can be inefficient.

The Dolev-Klawe-Rodeh algorithm~\cite{ring-elect-DKR} is an example of a variation of Chang and Roberts’ Algorithm. It improves efficiency by segmenting the ring and electing local leaders within each segment, reducing the total message complexity to $O(N \log N)$. This approach optimizes the election process by limiting the number of messages circulated. This is important because message complexity usually scales up with the size of the network.

The Hirschberg-Sinclair algorithm~\cite{ring-elect-HS} elects a leader in a bidirectional ring with minimal message overhead. Each node probes progressively larger segments of the topology to find the node with the highest ID, ultimately allowing the highest-ID node to declare itself as the leader. This method is more efficient than the Chang-Roberts algorithm, particularly in larger networks, and tolerates message delays. This makes it more suitable for adaptation to asynchronous environments. 

While convenient, the ring topology has several downsides: the overhead of executing a ring-construction algorithm prior to the election, strict limitations on message routing, and high sensitivity to failures, as the crash of a single node necessitates ring repair. Another strength of Hirschberg-Sinclair is that its concepts can extend to tree structures.

\subsection{Ring with Anonymous Nodes}
A classic result shows that no deterministic algorithm for leader election is viable if nodes do not start with distinct identities~\cite{angluin-impossibility}. This stems from the impossibility for any node to differentiate itself from the others. The intuition for the proof is that, in a symmetric configuration where all nodes start in the same state, it is possible for them to reach another symmetric configuration at every step.
Thus in cases where nodes are anonymous, leader election relies on  probabilistic methods to break the symmetry. These algorithms are synchronous, often requiring known network size and careful coordination.

Itai and Rodeh’s algorithm~\cite{ring-elect-IR} is the most prominent algorithm designed for electing a leader in anonymous rings of known size. Nodes send phased messages around the ring to produce distinguishable patterns, enabling a single node to emerge as the leader. As mentioned earlier, if the network size is unknown, achieving deterministic leader election is impossible in anonymous rings without additional assumptions, as nodes cannot break symmetry reliably.

\subsection{Algorithms which Tolerate Failures}
\label{subsec:ft-elections}

In realistic distributed systems, nodes may crash, and messages may be delayed or lost. Fault-tolerant leader election algorithms are designed to work in these unreliable conditions, often making fewer assumptions about topology.

The Bully algorithm~\cite{elect-bully} tolerates failures by assuming each node knows the IDs of all other nodes and can send messages directly to any node. If a node detects a failure in the leader, it initiates an election by sending "election" messages to all nodes with higher IDs. 
If a higher-ID node responds, the initiating node defers to that node.
If no higher-ID node responds, the initiating node declares itself the leader and informs all other nodes.
The Bully algorithm requires $O(N^2)$ messages in the worst case, but its resilience to node failures makes it suitable for fault-prone environments where nodes may be temporarily unavailable. However, in highly dynamic networks with frequent failures, this algorithm can incur high message costs.

Leader election algorithms must be chosen based on network requirements, including assumptions about failures and topology. Algorithms for reliable, failure-free networks, such as those for rings with or without unique IDs, can be more efficient and simpler. In contrast, fault-tolerant algorithms add resilience for unpredictable or unreliable conditions, albeit with higher message costs. Together, these methods provide adaptable strategies for leader election across diverse distributed environments.

%% file: dqc-framework.tex
In this section we provide a framework for distributed quantum systems.  We prescribe both 
the hardware and operational requirements needed to establish distributed computing.  The 
distributed computing set up for quantum systems has two broad classes of hardware 
requirements {\it viz} {\it (i)} nodes and {\it (ii)} communication channels.  

\noindent {\bf Nodes}:  The nodes should have both quantum and classical components built 
into them.  

\noindent {\it Quantum Components:} The primary quantum component is the quantum processor. 
A quantum processor is a collection of connected qubits. These qubits should be capable of co-operating with each other and perform computing tasks. These qubits, should be 
individually controllable and measurable. The relaxation time $T_{1}$ and the decoherence time $T_{2}$ of each of the individual qubits should be known in advance.  

\noindent {\it Classical components:} The classical components in the system are the 
classical registers capable of storing information using binary logic. These classical 
registers serve as classical memory and are used to store the information obtained from the 
quantum measurements on the qubits. Since we are working with distributed systems, and since
shared memory would limit the scalability of the system(as explained in Section~\ref{sec:overview}), 
each quantum processor should have their own classical memory. This implies that every quantum 
processor should have its own set of classical registers.

\noindent {\bf Channels}:  
The qubits should be capable of communicating with each other via channels.  In general for any quantum communication, we require quantum channels 
as well as classical channels.  The quantum channels should be capable of preserving the coherence of the qubits carrying quantum information during 
transmission.  

%% file: quantum-election-algo.tex
This section details our quantum leader election algorithm.

Let us consider a set of $n$ nodes, each with its own identifier. A node identifier is a sequence composed of $\lceil \log_{2} n \rceil$ binary digits.
Our algorithm relies on sharing sets of quantum states. All nodes use the same measurement basis for quantum states: in our case we assume that they use the $
\sigma^{z}$ computational basis. All nodes know in advance the default order in which they must measure the different states in a set. We also make the 
following assumptions: 

(a) We assume an arbitrary topology and a partially synchronous temporal model. 

(b) For the sake of simplicity, we start by assuming that nodes don't fail.

\begin{figure*}
    \centering
    \includegraphics[width=\textwidth]{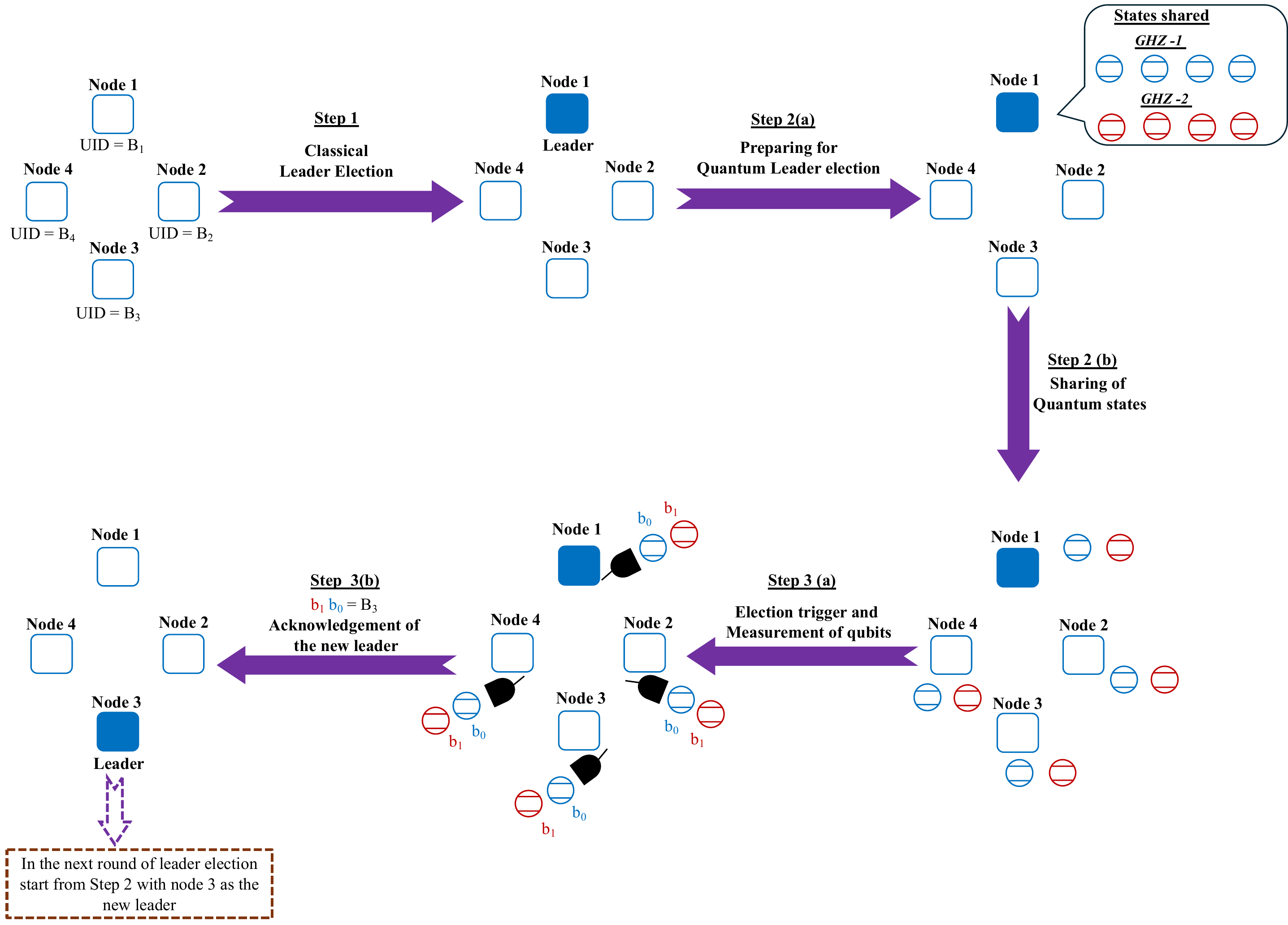}
    \caption{A schematic picture of the quantum leader election algorithm. }
    \label{fig:qleader-election}
\end{figure*}

Given the above assumptions, Figure~\ref{fig:qleader-election} details the steps of our leader election algorithm. It proceeds as follows: 

\begin{enumerate}
    \item \emph{Initial Election:} The first step involves electing a leader $L$ using a suitable classical leader election protocol. \emph{This step only occurs once.} Any subsequent election will start from the next step.

    \item \emph{Quantum Preparation:} The elected leader $L$ performs the following operations to prepare the system for a quantum leader election:
           \begin{enumerate}
               \item $L$ prepares a set $S_Q$ of $\lceil \log_{2} n \rceil$ quantum states to be shared between all the $N$ nodes in the system. Every quantum state in $S_Q$ is a multipartite entangled GHZ state of the form 
                       $(|0 \cdots 0 \rangle + |1 \cdots 1 \rangle)/\sqrt{2}$.

               \item $L$ shares $S_Q$ with all the other nodes. Upon reception, every node in the network has $\lceil \log_{2} n \rceil$ qubits, each coming from the $\lceil \log_{2} n \rceil$ different GHZ states.  
           \end{enumerate}

    \item \emph{Quantum Election:} To trigger a new election, a node sends a message $m_E$ to all nodes via a classic channel.

            \begin{enumerate}
                \item Upon reception of $m_E$, every node carries out a measurement on each of their $\lceil \log_{2} n \rceil$ qubits in the default order. The measurement output for each qubit is either a $0$ or a $1$; so the set of $\lceil \log_{2} n \rceil$ states gives a binary sequence $B$. 

                \item Each node compares their identifier with $B$. The new leader $L'$ is the node whose identifier matches $B$. 
                
                $L'$ waits for an acknowledgment message from all the other nodes, and then prepares the next quantum leader election by performing the operations of the \emph{Quantum Preparation} step.
                
                All the other nodes become leaderless, and send an acknowledgment message to $L'$. Then they wait for the new shared set of quantum states from $L'$. Upon reception of this new shared set, a node can switch to $L'$ as its new leader.
            \end{enumerate}
    
\end{enumerate}

The message complexity of this quantum election algorithm is $O(n)$: $n-1$ messages to trigger a reelection, $n-1$ messages to acknowledge the new leader, and $n-1$ messages to share the states for the next election. The time complexity is $O(1)$, since messages are sent concurrently.

Now let us assume that a maximum of $f$ node crash failures may occur simultaneously. A node that detects the failure of the current leader triggers  
a new election. The quantum states measurement produces binary sequence $B$. It is possible that $B$ matches the identifier of a crashed node. To solve 
this issue, all nodes whose identifier is in $[B, B+f]$ carry out the Bully algorithm (see Subsection~\ref{subsec:ft-elections}) among themselves.

The worst-case message complexity of our fault-tolerant quantum election algorithm is $O(n + f^2)$. However, since $f$ is typically much smaller than $n$, its contribution to the complexity is negligible. As a result, even in the presence of failures, the expected message complexity of our algorithm will likely remain $O(n)$.

%% file: blueprint.tex
Designing distributed quantum algorithms requires a structured approach that incorporates the realities of asynchronous networks, classical communication 
constraints, and partial synchrony models. In distributed quantum systems, quantum operations and entanglement can significantly reduce communication 
overhead. However, classical communication remains essential for setting up synchronization and coordination, and for handling failures. This section 
integrates these elements by focusing on two principles: compatibility with asynchronous assumptions about the network, and a balance between classical and 
quantum communication channels

\emph{Asynchronous Compatibility.} Quantum algorithms must be robust to delays and timing uncertainties inherent in asynchronous networks. Techniques like 
entanglement swapping and quantum error correction can help, but the design must assume that some message exchanges will occur out of sync. Partial synchrony 
assumptions, where network timing eventually falls within predictable bounds, provide a feasible middle ground. By designing for eventual synchronicity rather 
than constant synchronicity, the algorithm can capitalize on moments when timing constraints are met while maintaining fault tolerance during asynchronous 
periods.
    
\emph{Classical and Quantum Communication Balance.} While quantum channels enable entanglement, classical channels are still required for measurement results 
and control signals. Effective distributed quantum algorithms should minimize the dependence on classical messages, using quantum communication wherever 
possible but ensuring fallback mechanisms through classical channels to maintain consistency.

The blueprint we propose supports quantum-enhanced operations, like leader election, by maximizing the use of quantum resources (e.g., entanglement for 
snap consistency across nodes) and limiting classical communication to essential coordination steps.

\subsection{Design Considerations and Practical Constraints}

Implementing distributed quantum algorithms introduces practical challenges that must be addressed to ensure reliability and efficiency.

Quantum states are highly sensitive to noise, which leads to decoherence.  To overcome this quantum error correction schemes like the use of 
logical qubit (a set of qubits to represent one quantum state) instead of physical qubit (single qubit to denote a quantum state) or dual rail 
encoding can be used.  These schemes offer protection to the qubits from environmental interference particularly in the entanglement transfer 
and measurement steps.  These error correction schemes extend the life time of the qubits long enough so that the leader election process can be 
carried out before the qubits decohere.

In Section~\ref{sec:distr-quantum-election}, our protocol assumes that the identifiers are binary digits and we use
$\lceil n \rceil$ number of GHZ states to find the leader.  An alternative method would be to consider the identifiers to be decimal numbers 
from $0$ to $N-1$ and use a shared single $N$-level GHZ state $(|0 \cdots 0 \rangle + |1 \cdots 1 \rangle + \cdots + |(N-1) \cdots (N-1) \rangle)/\sqrt{N}$ 
between the nodes.  Upon election, all nodes carry out a measurement and the leader is the node whose identifier matches with the measurement 
output.  While this process reduces the number of quantum states needed for leader election, it might not be possible to use a single qudit when 
we have hundreds of nodes.  This can be overcome by using $\lceil \log_{d} n \rceil$ number of qudits, where $d$ is the dimension of the qudits 
and repeating the protocol in Sec. \ref{sec:distr-quantum-election}.  This shift from dimension $2$ to dimension $d$ will give us an advantage through a 
reduction in the number of quantum states we need to prepare for the leader election process.

To handle asynchronous delays, distributed quantum computing implementations can integrate building blocks from the field of distributed systems. First and foremost, \emph{failure detectors} allow to assume partial synchrony by introducing adaptive time bounds in fully asynchronous environments~\cite{unreliable-FDs, hierarchical-FD, accrual, stab-FD}. Additional algorithmic tools include \emph{checkpointing} and \emph{quorum-based decision-making}. Checkpointing techniques~\cite{checkpointing1, checkpointing2} allow nodes to periodically synchronize their states, and thus prevent restarting an entire distributed computation from scratch if a failure causes the loss of critical data. When the network incurs temporary partitioning or when some nodes are slow to respond, \emph{quorum-based consistency}~\cite{quorums1, quorums2} helps maintain reliable data consistency by requiring agreement from a subset of nodes (a quorum). All of these techniques help maintain coherence in the network and avoid complete reliance on synchronous operations.

These design considerations are critical for building distributed quantum algorithms that can operate over long distances and remain resilient in asynchronous environments.

\subsection{Case Study – Quantum Leader Election Algorithm}

To illustrate our blueprint, consider our quantum leader election algorithm presented in Section~\ref{sec:distr-quantum-election}. It targets a distributed quantum network, assumes an arbitrary topology and partial synchrony, and tolerates failures.

\begin{enumerate}
    \item \textit{Initialization}: Each node has a unique identifier, which it uses as a basis for the initial classical leader election. This classical phase elects a temporary leader to facilitate the setup of quantum resources across the network. This phase \emph{only occurs once}, at the very beginning, to set up a configuration where a leader can prepare the next round of election. All future rounds will benefit from the significant efficiency improvement gained through entangled states.
    
    \item \textit{Entanglement Preparation with GHZ States}: The temporary leader prepares a series of GHZ states, distributing one qubit from each GHZ state to each node in the network. This entanglement allows all nodes to be correlated, establishing a shared basis for comparison during leader election.
    
    \item \textit{Quantum Measurement and Leader Selection}: Each node measures its qubits in the predefined basis, generating a binary sequence derived from the GHZ states. This sequence serves as a "quantum signature" for each node, allowing nodes to compare identifiers and elect the node with the highest matching sequence as the new leader. This form of "\emph{snap consistency}" is a powerful outcome of using quantum states in a distributed system, as a single local operation at each node leads to a tremendous reduction in message complexity and time complexity. Its cost is limited since it involves a small number of nodes: $f + 1$, where $f$ is the maximum expected number of simultaneous failures.

    \item \textit{Resilience and Fault Tolerance}: In case of node failure, the algorithm initiates a reelection by having all nodes with the next highest identifiers rerun the classical election protocol in parallel. This fallback mechanism, inspired by the Bully algorithm, ensures that the system can adapt to failures dynamically.
\end{enumerate}

This quantum leader election algorithm demonstrates the use of both classical and quantum resources, with GHZ states enabling efficient coordination and fallback mechanisms ensuring robustness in case of failures.

\subsection{Analysis of Performance and Scalability}

To evaluate the effectiveness of distributed quantum algorithms, let's compare the performance benefits and limitations of quantum-enhanced leader election relative to classical methods.

As regards \emph{message complexity}, an entanglement setup phase can minimize classical communication rounds. Our quantum leader election algorithm has an expected message complexity of $O(n)$, where $n$ represents the number of nodes. This is a reduction from the $O(n^2)$ message complexity of most classical leader election algorithms.
    
With partial synchrony, \emph{time complexity} depends on moments when the system behaves synchronously. Locally on the leader, the entanglement phase requires time proportional to the number of GHZ states distributed. But once established, the election phase completes in $O(1)$ time, as measurements are performed simultaneously across nodes.
    
Quantum entanglement allows for high \emph{scalability}. Sharing qubits entirely removes latency concerns for the concurrent execution step they apply to. 
Combining several multipartite entanglements can increase the number of nodes involved in these concurrent execution steps on a very large scale. However, 
the reliance on classical channels for initial synchronization limits efficacy in highly asynchronous networks.  Further we also require a careful 
management of entanglement since we are working with large networks.

This analysis shows that distributed quantum algorithms can offer substantial improvements in efficiency and scalability. Yet they remain subject to classical network limitations, reinforcing the persistence of FLP impossibility under asynchronous constraints.